\documentclass[apj]{emulateapj}
\usepackage{natbib}

\newcommand{\vsini}{\mbox{$v \sin i$}}

\newcommand{\mactrb}{\mbox{$v_{\rm mac}$}}

\newcommand{\kms}{\mbox{km\,s$^{-1}$}}
\newcommand{\ms}{\mbox{m\,s$^{-1}$}}

\newcommand{\mjup}{\mbox{M$_{\rm Jup}$}}
\newcommand{\rjup}{\mbox{R$_{\rm Jup}$}}
\newcommand{\mstar}{\mbox{$M_{*}$}}
\newcommand{\rstar}{\mbox{$R_{*}$}}
\newcommand{\msol}{\mbox{$\,{\rm M}_\odot$}}
\newcommand{\rsol}{\mbox{$\,{\rm R}_\odot$}}

\begin{document}

\title{WASP-17\lowercase{b}: an ultra-low density planet in a probable
retrograde orbit\altaffilmark{$\dagger$}}

\shorttitle{WASP-17b: an ultra-low density planet}
\shortauthors{D.~R.~Anderson et al.}

\author{D.~R.~Anderson\altaffilmark{1}, 
C.~Hellier\altaffilmark{1}, 
M.~Gillon\altaffilmark{2,3}, 
A.~H.~M.~J.~Triaud\altaffilmark{2}, 
B.~Smalley\altaffilmark{1}, 
L.~Hebb\altaffilmark{4}, 
A.~Collier~Cameron\altaffilmark{4}, 
P.~F.~L.~Maxted\altaffilmark{1}, 
D.~Queloz\altaffilmark{2}, 
R.~G.~West\altaffilmark{5}, 
S.~J.~Bentley\altaffilmark{1}, 
B.~Enoch\altaffilmark{4}, 
K.~Horne\altaffilmark{4}, 
T.~A.~Lister\altaffilmark{6}, 
M.~Mayor\altaffilmark{2},
N.~R.~Parley\altaffilmark{4}, 
F.~Pepe\altaffilmark{2},
D.~Pollacco\altaffilmark{7}, 
D. S\'egransan\altaffilmark{2},
S. Udry\altaffilmark{2},
D.~M.~Wilson\altaffilmark{1}\altaffilmark{$\star$}}
\email{dra@astro.keele.ac.uk}

\altaffiltext{1}{Astrophysics Group, Keele University, Staffordshire, ST5 5BG, 
UK}
\altaffiltext{2}{Observatoire de Gen\`eve, Universit\'e de Gen\`eve, 51 Chemin 
des Maillettes, 1290 Sauverny, Switzerland}
\altaffiltext{3}{Institut d'Astrophysique et de G\'eophysique,  Universit\'e de 
Li\`ege,  All\'ee du 6 Ao\^ut, 17,  Bat.  B5C, Li\`ege 1, Belgium}
\altaffiltext{4}{School of Physics and Astronomy, University of St. Andrews, 
North Haugh, Fife, KY16 9SS, UK}
\altaffiltext{5}{Department of Physics and Astronomy, University of Leicester, 
Leicester, LE1 7RH, UK}
\altaffiltext{6}{Las Cumbres Observatory, 6740 Cortona Dr. Suite 102, Santa 
Barbara, CA 93117, USA}
\altaffiltext{7}{Astrophysics Research Centre, School of Mathematics \& Physics,
 Queen's University, University Road, Belfast, BT7 1NN, UK}
\altaffiltext{$\star$}{Present address: Centre for Astrophysics \& Planetary 
Science, University of Kent, Canterbury, Kent, CT2 7NH, UK}
\altaffiltext{$\dagger$}{Based in part on data collected with the HARPS 
spectrograph at ESO La Silla Observatory under programme ID 081.C-0388(A).}

\begin{abstract}
We report the discovery of the transiting giant planet WASP-17b,
the least-dense planet currently known. It is 1.6 Saturn
masses but 1.5--2 Jupiter radii, giving a density of 6--14
per cent that of Jupiter. WASP-17b is in a 3.7-day orbit around
a sub-solar metallicity, $V$ = 11.6, F6 star.  Preliminary
detection of the Rossiter--McLaughlin effect suggests that
WASP-17b is in a retrograde orbit ($\lambda \approx -150$\,deg),
indicative of a violent history involving planet--planet or
star--planet scattering.

WASP-17b's bloated radius could be due to tidal heating resulting
from recent or ongoing tidal circularisation of an eccentric orbit,
such as the highly eccentric orbits that typically result from
scattering interactions. It will thus be important to
determine more precisely the current orbital eccentricity by
further high-precision radial velocity measurements or by
timing the secondary eclipse, both to reduce the uncertainty
on the planet's radius and to test tidal-heating models.
Owing to its low surface gravity, WASP-17b's atmosphere has the
largest scale height of any known planet, making it a good
target for transmission spectroscopy.
\end{abstract}


\section{Introduction}
The first measurement of the radius and density of an extrasolar planet was made
when HD\,209458b was seen to transit its parent star 
(Charbonneau et al. 2000, Henry et al. 2000).
The large radius (1.32\,\rjup) of HD\,209458b, confirmed by later observations 
(e.g., Knutson et al. 2007), could not be explained  by standard models of 
planet evolution (Guillot \& Showman 2002). 
Since the discovery of HD\,209458b, other bloated planets have been 
found, including 
TrES-4 (Mandushev et al. 2007), 
WASP-12b (Hebb et al. 2008), 
WASP-4b (Wilson et al. 2008; Gillon et al. 2009a, Winn et al. 2009a, Southworth
et al. 2009), 
WASP-6b (Gillon et al. 2009b), 
XO-3b (Johns-Krull et al. 2008; Winn et al. 2008) and 
HAT-P-1b (Bakos et al. 2007; Winn et al. 2007; Johnson et al. 2008). 
Of those, TrES-4 is the most bloated, with a density 15 per cent that of 
Jupiter, and a radius larger by a factor 1.78 (Sozzetti et al. 2009).

The mass, composition and evolution history of a planet determines its current
radius (e.g., Burrows et al. 2007; Fortney et al. 2007). 
Recently, numerous theoretical studies have attempted to discover the 
reasons why some short-orbit, giant planets are bloated. 
A small fraction of stellar insolation energy would be sufficient to 
account for bloating, but no known mechanism is able to transport the insolation 
energy deep enough within a planet to significantly affect the planet's 
evolution (Guillot \& Showman 2002; Burrows et al. 2007).
Enhanced atmospheric opacity would cause internal heat to be lost more slowly,
causing a planet's radius to be larger than otherwise at a given age 
(Burrows et al. 2007). 
Indeed, the more highly irradiated planets are thought to have enhanced
opacity due to species such as gas-phase TiO/VO, tholins or polyacetylenes
(Burrows et al. 2008; Fortney et al. 2008).
These upper-atmosphere absorbers result in detectable stratospheres 
(e.g., Knutson et al. 2009) and
prevent incident flux from reaching deep into the atmosphere, causing a large 
day-night temperature contrast, which leads to faster cooling (Guillot \& 
Showman 2002). 
That some planets are not bloated, though they are in similar irradiation 
environments and have otherwise similar properties to bloated planets, may be 
due to differences in evolution history or in core mass (Guillot et al. 2006; 
Burrows et al. 2007).

Currently, the most promising explanation for the large radii of some planets is
that they were inflated when the tidal circularisation of eccentric orbits 
caused energy to be dissipated as heat within the planets 
(Bodenheimer et al. 2001; Gu et al. 2003; 
Jackson et al. 2008a; Ibgui \& Burrows 2009). 
Indeed, Jackson et al. (2008b) found that the distribution of the eccentricities 
of  short-orbit ($a < 0.2$ AU) planets could have evolved, via tidal
circularisation, from a distribution identical to that of the farther-out 
planets.

The angular momenta of a star and its planets derive from that of their parent  
molecular cloud, so close alignment is expected between the stellar spin 
and planetary orbit axes.
When a planet obscures a portion of its parent star we observe an apparent 
spectroscopic redshift or blueshift; which we see depends on whether the area 
obscured is approaching or receding relative to the star's bulk motion.
This manifests as an `anomalous' radial velocity (RV) and is known as the
Rossiter-McLaughlin (RM) effect (e.g., Queloz et al. 2000a; Gaudi \& Winn 2007). 
The shape of the RM effect is sensitive to the path a planet takes across its
parent star, relative to the star's spin axis.
Thus, spectroscopic observation of a transit allows measurement of 
$\lambda$, the sky-projected angle between the stellar spin and planetary orbit 
axes. 
Short-orbit, giant planets are thought to have formed just outside the ice
boundary and migrated inwards (e.g., Ida \& Lin 2004).
Thus, $\lambda$ is a useful diagnostic for theories of planet migration, some of
which predict preservation of initial spin-orbit alignment and some of which 
would occassionally produce large misalignments.
For example, migration via tidal interaction of a giant planet with a gas disc 
(Lin et al. 1996; Ward 1997) is expected to preserve spin-orbit alignment,
whereas migration via a combination of planet-planet scattering and tidal 
circularisation of a resultant eccentric orbit is able to produce a
significant misalignment (e.g., Rasio \& Ford 1996; Chatterjee et al. 2008; 
Nagasawa  et al. 2008).
To date, $\lambda$ has been determined for 14 systems (Fabrycky \& Winn 2009; 
Gillon 2009; Triaud et al. 2009) 
and for 3 of those a significant misalignment was found: XO-3b 
($\lambda = 37.3 \pm 3.7$~deg, Winn et al. 2009b; see also: H\'ebrard et al. 
2008), HD\,80606b ($\lambda = 59^{+28}_{-18}$~deg, Gillon 2009; see also: 
Moutou et al. 2009; Pont et al. 2009; Winn et al. 2009c) 
and WASP-14b ($\lambda = -33.1 \pm 7.4$~deg, Johnson et al. 2009; see also: 
Joshi et al. 2009).

In this paper, we present the discovery of the transiting extrasolar 
planet WASP-17b, which is the least-dense planet currently known and 
the first planet found to be in a probable retrograde orbit.

\section{Observations}
\begin{figure}
\plotone{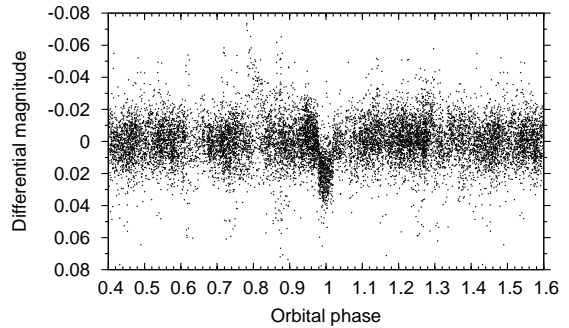}
\caption{WASP-South discovery light curve, phase-folded with the ephemeris of
Table~\ref{sys-params2}.
Points with error $>0.05$\,mag (3\,$\sigma_{median}$) were clipped for display. 
From the WASP discovery photometry we found a high probability (0.74) of WASP-17
being a main-sequence star and a zero probability of the companion having 
$R_{\rm P} < 1.5\,\rjup$ (Collier Cameron et al. 2007). 
As such, the system did not fulfill one of our 
usual selection criteria, $\mathcal{P}(R_{P}<1.5\,\rjup)>0.2$, 
for follow-up spectroscopy (Collier Cameron et al. 2007). We therefore advise 
other transit surveys to exercise caution in rejecting candidates on the basis 
of size, so as not to miss interesting systems like WASP-17.
\label{figWASP}}
\end{figure}

WASP-17 is a $V = 11.6$, F6 star in Scorpius. 
It was observed by WASP-South (Pollacco et al. 2006) from 2006 May 04 
to 2006 August 18, again from 2007 March 05 to 2007 August 19 and again from
2008 March 02 to 2008 April 19.
These observations resulted in 15\,509 usable photometric measurements, spanning
two years and from two separate fields.
A transit search (Collier Cameron et al. 2006) found a strong, 3.7-day 
periodicity (Figure~\ref{figWASP}).

\begin{figure}
\plotone{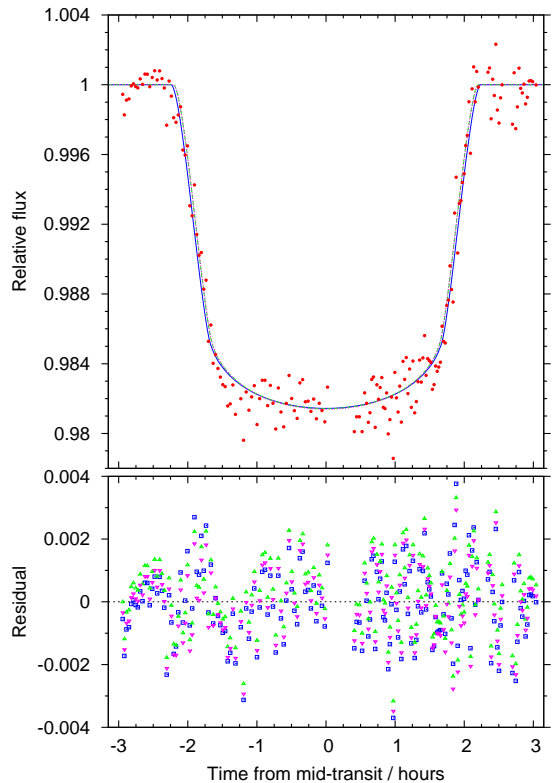}
\caption{{\bf Upper panel}: Euler $I_{c}$-band light curve (red circles) taken 
on 2008 May 06.
Overplotted are the best-fitting model transits (solid lines; 
\S\ref{sec:sysparams}) from the parameters of Table~\ref{sys-params2}; 
consult that table for the key to the colour and symbol scheme. 
{\bf Lower panel}: Residuals about the model fits.
For the green and the magenta models (triangles) the noise is the same 
(rms = 1140\,ppm; red noise = 840\,ppm --- calculated using the method of Gillon
et al. 2006); 
for the blue model (squares) the noise is slightly higher 
(rms = 1210\,ppm; red noise = 920\,ppm, Gillon et al. 2006). 
The mean theoretical error is 800\,ppm.
\label{figEuler}}
\end{figure}

A full transit of WASP-17 was observed in the $I_{c}$-band with EulerCAM on the 
1.2-m Euler-Swiss telescope on 2008 May 06.
The telescope was defocused to give a mean stellar profile width of 4\,\arcsec.
Over a duration of 6 hours, 181 frames were obtained with a range of exposure 
times of 32--98\,s --- the exposure time was tuned to keep the stellar peaks
constant. 
Observations began when WASP-17 was at airmass 1.07; the star then passed 
through the meridian before reaching airmass 1.8 when observations ended at
twilight. 
The resulting light curve and the residuals about the model fits 
(\S\ref{sec:sysparams}) are shown in Figure~\ref{figEuler} 
and the photometry is given in Table~\ref{tabEuler}.

\begin{deluxetable}{ccc}
\tablewidth{15pc}
\tablecaption{$I_{c}$-band photometry of WASP-17 \label{tabEuler}} 
\tablehead{
  \colhead{HJD--2\,450\,000} & 
  \colhead{Relative flux} & 
  \colhead{$\sigma_{flux}$} \\
  \colhead{(days)} & 
  \colhead{} &
  \colhead{}
}
\startdata
4592.677426 & 0.999803  & 0.000959 \\
4592.678457 & 0.998623  & 0.000956 \\
4592.679703 & 0.999471  & 0.000670 \\
\ldots      & \ldots    & \ldots   \\
4592.923564 & 1.00053~  & 0.00141~ \\
4592.926643 & 1.00034~  & 0.00145~
\enddata
\tablecomments{This table is presented in its entirety in the electronic 
edition of the Astrophysical Journal. A portion is shown here for guidance
regarding its form and content.}
\end{deluxetable}

\begin{figure}
\plotone{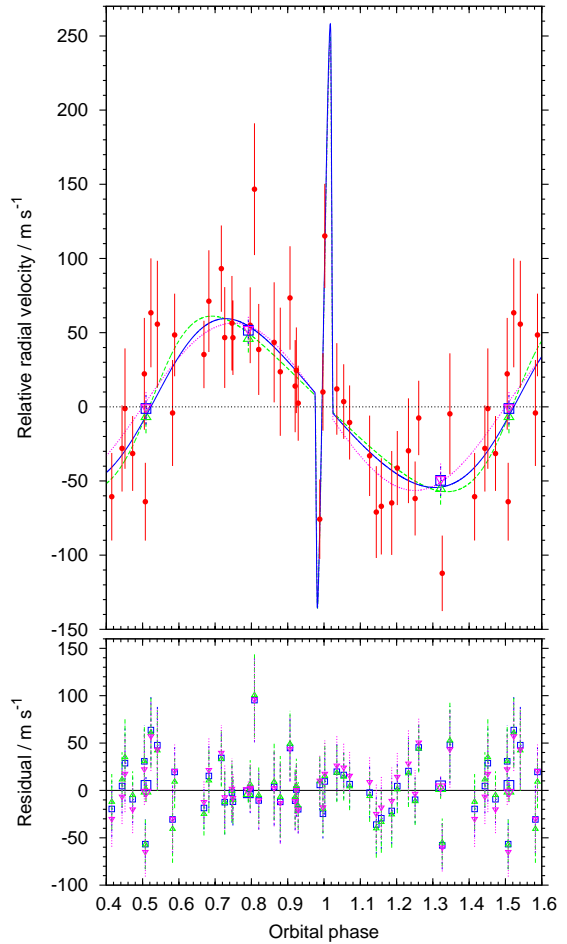}
\caption{{\bf Upper panel}: Relative radial velocity measurements of WASP-17 as 
measured by CORALIE (red circles).
The three, colour-coded solid lines are the model orbital solutions 
(\S\ref{sec:sysparams}) based 
on the parameters of Table~\ref{sys-params2} and incorporate the RM effect.
As the zero-point offset between HARPS and CORALIE is a free parameter in the 
models, the HARPS measurements are shown once per model, with corresponding 
symbols and colours (Table~\ref{sys-params2}).
The centre-of-mass velocities of Table~\ref{sys-params2} have been subtracted. 
{\bf Lower panel}: Residuals about the model solutions; consult 
Table~\ref{sys-params2} for the key to the symbol and colour scheme.
\label{figRV}}
\end{figure}

Using the CORALIE spectrograph mounted on the Euler-Swiss telescope 
(Baranne et al. 1996; Queloz et al. 2000b), 
5 spectra of WASP-17 were obtained in 2007, 16 more in 2008, and a further 
20 in 2009.
Three high-precision spectra were obtained 
in 2008 with the HARPS spectrograph (Mayor et al. 2003), based on the 3.6-m ESO 
telescope. 
RV measurements were computed by weighted 
cross-correlation (Baranne et al. 1996; Pepe et al. 2005) 
with a numerical G2-spectral template. 
RV variations were detected with the same period found from the WASP photometry 
and with semi-amplitude of $\sim$50\,\ms, consistent with a 
planetary-mass companion.
The RV measurements are listed in Table~\ref{rv-data} and are plotted in
Figure~\ref{figRV}.

\begin{figure}
\plotone{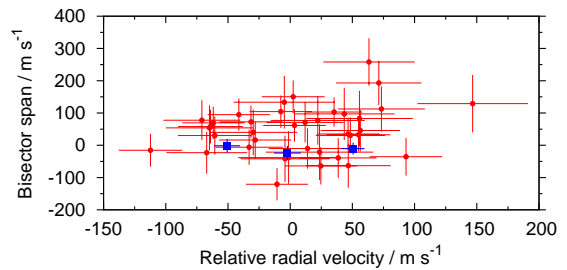}
\caption{Bisector span versus relative radial velocity for the CORALIE (red
circles) and HARPS (blue squares) spectra. 
Averages of the centre-of-mass velocities and zero-point offsets of
Table~\ref{sys-params2} were subtracted. 
Bisector uncertainties equal to twice the radial velocity uncertainties have 
been adopted.
The Pearson correlation coefficient is 0.19.\label{figBis}}
\end{figure}

To test the hypothesis that the RV variations are due to spectral line 
distortions caused by a blended eclipsing binary, a line-bisector analysis 
(Queloz et al. 2001) of the CORALIE and HARPS cross-correlation functions was 
performed. 
The lack of correlation between bisector span and radial velocity 
(Figure~\ref{figBis}), especially for the high precision HARPS measurements, 
supports the identification of the transiting body as a planet.

\begin{table}
\centering
\caption{Radial velocity measurements of WASP-17
\label{rv-data}} 
\begin{tabular}{cccc}
\hline
\hline
BJD--2\,450\,000 & RV & $\sigma$$_{RV}$ & BS$^{a}$\\ 
  & (\kms) & (\kms) & (\kms)\\ 
\hline
\multicolumn{4}{l}{CORALIE:}\\
4329.6037 & --49.4570 & 0.0428 & ~\,0.0829\\
4360.4863 & --49.3661 & 0.0444 & ~\,0.1290\\
4362.4980 & --49.5175 & 0.0407 & ~\,0.1335\\
4364.4880 & --49.4891 & 0.0432 & --0.0213\\
4367.4883 & --49.4415 & 0.0343 & ~\,0.1931\\
4558.8839 & --49.4988 & 0.0311 & --0.0101\\
4559.7708 & --49.5798 & 0.0325 & --0.0229\\
4560.7314 & --49.5734 & 0.0295 & ~\,0.0303\\
4588.7799 & --49.4881 & 0.0289 & --0.0638\\
4591.7778 & --49.4661 & 0.0340 & --0.0631\\
4622.6917 & --49.3976 & 0.0351 & ~\,0.1071\\
4624.6367 & --49.4494 & 0.0367 & ~\,0.2581\\
4651.6195 & --49.4564 & 0.0319 & ~\,0.0459\\
4659.5246 & --49.4693 & 0.0405 & ~\,0.0972\\
4664.6425 & --49.5424 & 0.0353 & ~\,0.0401\\
4665.6593 & --49.4905 & 0.0377 & ~\,0.0770\\
4682.5824 & --49.5007 & 0.0309 & ~\,0.0707\\
4684.6264 & --49.5169 & 0.0357 & --0.0418\\
4685.5145 & --49.4741 & 0.0307 & --0.0393\\
4690.6182 & --49.5776 & 0.0350 & ~\,0.0541\\
4691.6077 & --49.5140 & 0.0406 & --0.0380\\
4939.8457 & --49.4394 & 0.0349 & ~\,0.1127\\
4940.7346 & --49.5838 & 0.0309 & ~\,0.0777\\
4941.8520 & --49.5408 & 0.0290 & ~\,0.0163\\
4942.6959 & --49.4775 & 0.0227 & ~\,0.1030\\
4942.8747 & --49.4196 & 0.0291 & --0.0355\\
4943.6655 & --49.5103 & 0.0253 & ~\,0.1504\\
4943.8872 & --49.5885 & 0.0268 & ~\,0.0732\\
4944.6858 & --49.5540 & 0.0250 & ~\,0.0951\\
4944.8689 & --49.5746 & 0.0249 & ~\,0.0694\\
4945.6969 & --49.5442 & 0.0252 & ~\,0.0727\\
4945.8277 & --49.5767 & 0.0263 & ~\,0.0579\\
4946.7289 & --49.4662 & 0.0251 & ~\,0.0351\\
4946.9069 & --49.4578 & 0.0256 & ~\,0.0326\\
4947.6558 & --49.5028 & 0.0261 & --0.1128\\
4947.8694 & --49.5092 & 0.0251 & ~\,0.0614\\
4948.6415 & --49.5203 & 0.0251 & ~\,0.1043\\
4948.8836 & --49.6250 & 0.0254 & --0.0153\\
4949.8646 & --49.4643 & 0.0279 & ~\,0.0307\\
4951.6661 & --49.5233 & 0.0250 & --0.1208\\
4951.8719 & --49.5458 & 0.0271 & --0.0061\\
\multicolumn{4}{l}{HARPS:}\\
4564.8195 & --49.4884 & 0.0108 & --0.0239\\
4565.8731 & --49.4356 & 0.0092 & --0.0103\\
4567.8516 & --49.5368 & 0.0105 & --0.0015\\
\hline
\multicolumn{4}{l}{$^{a}$ BS: bisector span}
\end{tabular} 
\end{table} 

\section{Stellar parameters}
The combined CORALIE and HARPS spectra from 2007--2008, 
co-added into 0.01\AA\ steps, give a S/N of $\sim$100:1.
The stellar parameters and elemental abundances of WASP-17 were determined
using spectrum synthesis and equivalent-width measurements (Gillon et al.
2009b; West et al. 2009) and are given in
Table~\ref{wasp17-params}. In the spectra, the Li {\sc i} 6708\AA\ line is not
detected (EW $<$ 2m\AA), giving an upper-limit on the Lithium
abundance of log n(Li/H) + 12 $<$ 1.3. However, the effective temperature of 
this star implies it is in the lithium-gap (B\"{o}hm-Vitense 2004) and so the 
lithium abundance does not provide an age constraint.
In determining the projected stellar rotation velocity (\vsini) 
from the HARPS spectra, a value for macroturbulence (\mactrb) of 6~\kms\ was 
adopted (Gray 2008) and an instrumental FWHM of 0.06\AA, determined from the 
telluric lines around 6300\AA, was used. 
A best fitting value of \vsini\ = 9.0 $\pm$ 1.5~\kms\ was
obtained. However, if \mactrb\ is lower than the assumed 6~\kms\, then \vsini\
would be slightly higher, with a value of 11~\kms\ obtained if \mactrb\ is
assumed to be zero.

We attempted to measure the rotation period of WASP-17 by searching for
sinusoidal, rotational modulation of the WASP light curve (Hebb et al. 2009), 
as may be induced by a non-axisymmetric distribution of starspots. 
Considering periods of 1.05--30\,days, the best-fitting period is 
$24.7$\,days. However, the amplitudes of the phase-folded light curves from
each camera from each season are small (2--8\,mmag).
Assuming spin-orbit alignment, with \vsini\ = 9.0\,\kms, and using the values of
stellar radius given in Table~\ref{sys-params2} (see \S\ref{sec:sysparams}), a 
stellar rotation period of 8.5--11\,days is expected.

We estimated the distance of WASP-17 ($400 \pm 60$~pc) using the distance
modulus, the TYCHO apparent visual magnitude ($V = 11.6$) and the absolute
visual magnitude of an F6V star ($V = 3.6$; Gray 2008);
we assumed $E(B-V)=0$.

\begin{table}[!h]
\centering
\caption{Stellar parameters of WASP-17} 
\begin{tabular}{lc}
\hline
\hline
Parameter & Value\\
\hline
$T_{\rm eff}$ (K)	& 6550 $\pm$ 100\\ 
$\log g_{*}$ (cgs)	& 4.2  $\pm$ 0.2\\ 
$\xi_{\rm t}$ (\kms)	& 1.6  $\pm$ 0.2\\
$v\sin i$ (\kms)   	& 9.0  $\pm$ 1.5\\
 [2mm]
Spectral Type		& F6		\\
  [2mm]
{[Na/H]}	   &$-$0.15 $\pm$ 0.06  \\
{[Mg/H]}	   &$-$0.21 $\pm$ 0.07  \\
{[Al/H]}	   &$-$0.38 $\pm$ 0.05  \\
{[Si/H]}	   &$-$0.18 $\pm$ 0.09  \\
{[Ca/H]}	   &$-$0.08 $\pm$ 0.14  \\
{[Sc/H]}	   &$-$0.20 $\pm$ 0.14  \\
{[Ti/H]}	   &$-$0.20 $\pm$ 0.12  \\
{[V/H]}		   &$-$0.38 $\pm$ 0.13  \\
{[Cr/H]}	   &$-$0.23 $\pm$ 0.14  \\
{[Fe/H]}	   &$-$0.25 $\pm$ 0.09  \\
{[Ni/H]}	   &$-$0.32 $\pm$ 0.11  \\
$\log N({\rm Li})$ &$<$ 1.3		\\
  [2mm]
  $V$ (mag)	   & 11.6		\\
Distance (pc)	   & 400    $\pm$ 60	\\
\hline
\multicolumn{2}{c}{R.A. (J2000) = 15$\rm^{h}$59$\rm^{m}$50.94$\rm^{s}$}\\
\multicolumn{2}{c}{Dec. (J2000) = --28$^\circ$03$^{'}$42.3$^{''}$} \\ 
\multicolumn{2}{c}{1SWASP J155950.94$-$280342.3}\\
\multicolumn{2}{c}{USNO-B1.0 0619-0419495}\\
\multicolumn{2}{c}{2MASS 15595095$-$2803422}\\
\hline
\end{tabular}
\label{wasp17-params}
\end{table}


\section{System parameters}
\label{sec:sysparams}
The WASP-South and EulerCAM photometry were combined with the CORALIE and HARPS
RV measurements in a simultaneous Markov-chain Monte-Carlo (MCMC) analysis 
(Collier Cameron et al. 2007; Pollacco et al. 2008). 
The proposal parameters we use are: $T_{\rm c}$, $P$, $\Delta F$, $T_{14}$, $b$,
$K_{\rm 1}$, $M_{*}$, $e \cos \omega$, $e \sin \omega$, $\vsini \cos \lambda$ 
and $\vsini \sin \lambda$. 
Here $T_{\rm c}$ is the epoch of mid-transit, $P$ is the orbital period, $\Delta
F$ is the fractional flux-deficit that would be observed during transit in the
absence of limb-darkening, $T_{14}$ is the total transit duration (from first to
fourth contact), $b$ is the impact parameter of the planet's path across the
stellar disc, $K_{\rm 1}$ is the stellar reflex velocity semi-amplitude, 
\mstar\ is the stellar mass, $e$ is the orbital eccentricity and $\omega$ is the
argument of periastron. 

At each step in the MCMC procedure, each proposal parameter is perturbed from
its previous value by a small, random amount. 
From the proposal parameters, model light and RV curves are generated and 
$\chi^{2}$ is calculated from their comparison with the data. 
A step is accepted if $\chi^{2}$ is lower than for the previous step;
a step with higher $\chi^{2}$ may also be accepted, the
probability for which is lower for larger $\Delta \chi^{2}$. 
In this way, the parameter space around the optimum solution is thoroughly 
explored.
Provided the probability of accepting a step of higher $\chi^{2}$ is chosen
correctly, then the distribution of points for an MCMC chain gives the standard
errors on the parameters (e.g., Ford 2006).

\begin{figure}[!h]
\plotone{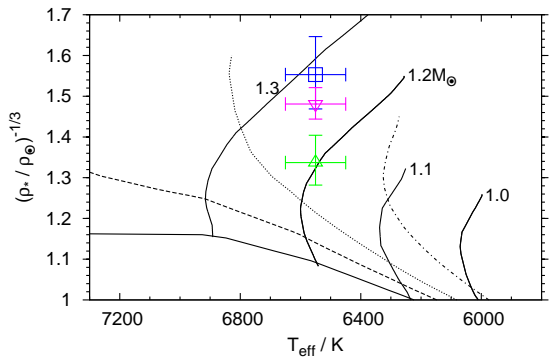}
\caption{Modified H-R diagram: 
inverse cube-root of stellar density versus 
stellar effective temperature. 
The former quantity is purely observational, measured 
directly from the light curve in initial MCMC runs. 
The latter quantity is derived from the spectral analysis. 
The best-fitting values of the three models for WASP-17 are depicted using the 
key of Table~\ref{sys-params2}. 
Evolutionary mass tracks (solid, labelled lines) and isochrones (100\,Myr,
solid; 1\,Gyr, dashed; 2\,Gyr, dotted; 4\,Gyr, dot-dashed) from Girardi et
al. (2000) for $[{\rm Fe/H}]=-0.25$ are plotted for comparison.
\label{figEvol}}
\end{figure}

We place a prior on \mstar\ that, via a Bayesian penalty, causes its values 
in accepted MCMC steps to approximate a Gaussian distribution with mean $M_{0}$ 
and standard deviation $\sigma_{\rm M} = 0.1\,M_{0}$, where $M_{0}$ is the 
initial estimate of \mstar.
To determine $M_{0}$ and to estimate the star's age, an evolutionary analysis 
(Hebb et al. 2008) was performed.
In that, an initial MCMC run was used to determine the stellar density, 
which depends on the shape of the transit light curve and the eccentricity of 
the orbit.
The stellar evolution tracks of Girardi et al. (2000) were then interpolated 
using this stellar density and using the stellar temperature and metallicity 
from the spectral analysis (Figure~\ref{figEvol}).
This suggests that WASP-17 has evolved off the zero-age main sequence, with 
a mass of $1.20^{+0.10}_{-0.11}$\msol\ and an age of $3.0^{+0.9}_{-2.6}$\,Gyr.
This stellar mass was used as the initial estimate, $M_{0}$, in the MCMC
solution (Case 1) presented in the second column of Table~\ref{sys-params2}. 
The best-fitting eccentricity is non-zero at the 2-$\sigma$ level 
($e=0.129^{+0.106}_{-0.068}$; $\omega = 290^{+106}_{-16}$\,deg). 
The best-fitting planet radius is large but uncertain 
($R_{\rm P} = 1.74^{+0.26}_{-0.23}\,\rjup$). 
This uncertainty results from $e$ and $\omega$ being poorly constrained by the 
RV data, causing the velocity of the planet during transit, and therefore the 
distance travelled (i.e. the stellar radius), to be uncertain. 
The planet radius is related to the stellar radius by the measured depth of
transit, so it too is uncertain.

The stellar age from the first MCMC solution and \vsini\ 
from the spectral analysis are consistent with WASP-17 being young. 
As such, a second MCMC analysis was performed, with a main sequence (MS) prior 
on the star and with eccentricity a free parameter.
With the MS prior, a Bayesian penalty ensures that, in accepted steps, the 
values of stellar radius are consistent with those of stellar mass for a 
main-sequence star: the probability distribution of \rstar\ 
has a mean of $R_{0} = M_{0}^{0.8}$ (Tingley \& Sackett 2005), where $R_{0}$ is 
the initial estimate of 
\rstar, and a standard deviation of $\sigma_{R} = 0.8(R_{0}/M_{0})\sigma_{M}$.
An initial MCMC analysis was used to determine stellar density, which was used 
as an input to an evolutionary analysis (Figure~\ref{figEvol}). 
That suggests a stellar mass of $1.19^{+0.07}_{-0.08}$\msol\ and a stellar age 
of $1.2^{+2.8}_{-1.2}$\,Gyr.
This stellar mass was used as the start value in the MCMC solution (Case 2) 
presented in the third column of Table~\ref{sys-params2}. 
The MS prior results in a smaller stellar radius and, therefore, a smaller 
planetary radius ($R_{\rm P} = 1.51 \pm 0.10\,\rjup$).
The MS prior on stellar radius, together with the prior on stellar mass,
effectively places a prior on stellar density, $\rho_{\rm *}$, forcing stellar
density toward the higher values typical of a MS star.
Therefore, as 
\begin{equation}
  \rho_{\rm *} \propto \frac{(1-e^{2})^{3/2}}{(1 + e \sin \omega)^{3}}
\end{equation}
a more eccentric orbit 
($e=0.237^{+0.068}_{-0.069}$; $\omega = 278.0^{+8.2}_{-5.6}$\,deg) results.
The uncertainties on each of the parameters affected by the MS prior are 
artificially small due to the MS prior not taking full account 
of uncertainties involved (e.g., in the theoretical mass-radius relationship).

As the detection of a non-zero eccentricity in the first MCMC solution is of low
significance, a third solution was generated, with an imposed circular orbit and
no MS prior. 
Again, an initial MCMC analysis was performed to determine stellar density, which
was used as an input to an evolutionary analysis (Figure~\ref{figEvol}).
From that, a stellar mass of $1.25 \pm 0.08$\msol\ and a stellar age of 
$3.1^{+1.1}_{-0.8}$\,Gyr was found.
This stellar mass was used as the start value in the MCMC solution (Case 3) 
presented in the fourth column of Table~\ref{sys-params2}. 
The circular orbit causes the velocity of the planet during transit to be 
higher than in the two eccentric solutions. 
This results in a larger stellar radius and, as the depth of transit is fixed
by measurement, in a larger planet radius 
($R_{\rm P} = 1.97 \pm 0.10\,\rjup$).

For each model, the best-fitting transit light curve is shown in 
Figure~\ref{figEuler} and the best-fitting RV curve is shown in 
Figure~\ref{figRV}.
To help decide between the three cases presented, a more
precise determination of stellar age, stellar radius or orbtial eccentricity
would be useful. 
It is currently difficult to reliably determine the age of stars older than
1--2\,Gyr (e.g., Sozzetti et al. 2009 and references therein).
Stellar radius could be calculated from a precise parallax determination.
WASP-17's parallax is predicted to be 2.5\,mas, which will be measurable to good
precision by the forthcoming Gaia mission (Jordi et al. 2006), which is expected
to achieve an accuracy of 7\,$\mu$as at $V = 10$. 
Eccentricity can be better determined using a combination of two methods:
(i) Take a number of high-precision RV measurements (which best constrain
$e \sin \omega$), focusing on those phases at which the differences between the 
models are greatest (Figure~\ref{figRV}).
(ii) Observe the secondary eclipse; 
the time of mid-eclipse constrains $e \cos \omega$ and the eclipse duration more
weakly constrains $e \sin \omega$ (Charbonneau et al. 2005).

We adopt Case 1 as our preferred solution; we note that if WASP-17 proves to be
young then Case 2 will be indicated, and Case 3 will be indicated if the
planet's orbit is found to be (near-)circular.

\subsection{A retrograde orbit?}
\begin{figure}[!h]
\plotone{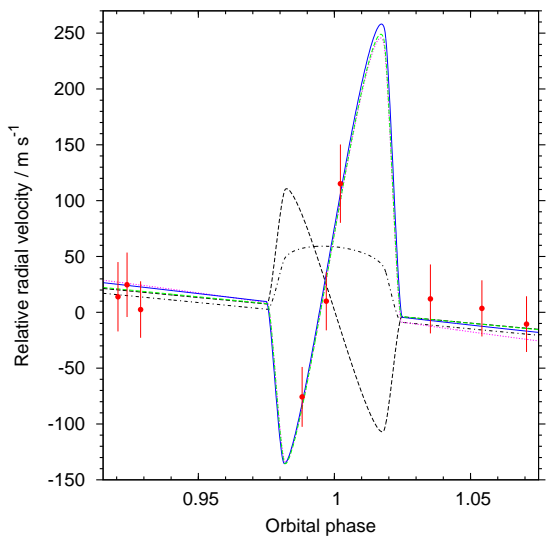}
\caption{Zoom-in on the spectroscopic transit region of 
Figure~\ref{figRV}, showing the fitted RM effect more clearly.
The red circles are CORALIE measurements and the coloured lines are the 
best-fitting models (Table~\ref{sys-params2}).
We show for comparison the RM effects that would result from perpendicular 
($\lambda = -90$\,deg; black, dot-dashed line) and aligned ($\lambda = 0$\,deg;
black, dashed line) spin-orbit axes; in both cases $\vsini\ = 9.0$\,\kms\ and 
$b = 0.355$ were fixed.
\label{figRM}}
\end{figure}

The RM effect was incorporated in the
MCMC analyses with free parameters $\vsini \cos \lambda$ and 
$\vsini \sin \lambda$ (Figure~\ref{figRM}; Table~\ref{sys-params2}).
The three RV measurements during transit suggest a large
spin-orbit misalignment ($\lambda \approx -150$\,deg), 
indicating that the planet is orbiting in a sense counter to that of stellar 
rotation. 
The RV RMS about the fitted model is 31.6\,\ms.
Comparing a borderline prograde-retrograde orbit 
($\lambda = -90$\,deg, $\vsini = 9.0\,\kms$, $b = 0.355$), 
the first in-transit point is discrepant by 5.0\,$\sigma$ and the RV RMS is 
40.2\,\ms\ (Figure~\ref{figRM}).
Comparing a prograde orbit ($\lambda = 0$\,deg, $\vsini\ = 9.0\,\kms$,
$b = 0.355$), the first and third in-transit points are discrepant by 
6.0\,$\sigma$ and 3.8\,$\sigma$ respectively, and the RV RMS is 45.7\,\ms\
(Figure~\ref{figRM}).

The fitted amplitude of the RM effect suggests $\vsini \approx 20\,\kms$, which
is higher than determined in the spectral analysis (Table~\ref{wasp17-params}).
This could be because the amplitude of the RM effect is currently liable to be 
overestimated (Winn et al. 2005; Triaud et al. 2009), due to the manner in which
the RVs are extracted from the spectra.
At present, the effective velocity of the spectroscopic 
cross-correlation function (CCF) is measured by fitting a gaussian. 
However, at values of \vsini\ significantly
greater than the intrinsic width of the CCF for a slowly-rotating star,
the travelling bump in the profile that is the spectral signature of the
planet's silhouette becomes partially resolved (Gaudi \& Winn 2007). 
The CCF profile will become
slightly asymmetric when the planet is near the limb, and this may bias
the velocity measured by gaussian-fitting to a greater value than the
RV of the centroid of the unobscured parts of the star.

\begin{table*}[!h]
\centering
\caption{System parameters of WASP-17} 
\label{sys-params2} 
\begin{tabular}{lccc} 
\hline 
\hline
Parameter 
& Case 1 (adopted)
& Case 2
& Case 3\\
\hline 
\\
Extra constraints 
& \ldots
& MS prior 
& $e=0$\\
\noalign{\smallskip}
Graph colour
& blue
& green
& magenta\\
\noalign{\smallskip}
Graph symbol
& squares
& upwards triangles
& downwards triangles\\
\noalign{\smallskip}
Graph line
& solid
& dashed
& dotted\\
\noalign{\smallskip}
Stellar age (Gyr)
& $3.0^{+0.9}_{-2.6}$
& $1.2^{+2.8}_{-1.2}$
& $3.1^{+1.1}_{-0.8}$\vspace{3 mm}\\
\noalign{\smallskip}
$P$ (d)
& 3.7354417$^{+0.0000072}_{-0.0000073}$ 
& 3.7354417$^{+0.0000073}_{-0.0000074}$
& 3.7354414$^{+0.0000074}_{-0.0000074}$\\
\noalign{\smallskip}
$T_{\rm c}$ (HJD) 
& 2454559.18102$^{+0.00028}_{-0.00028}$ 
& 2454559.18100$^{+0.00027}_{-0.00028}$ 
& 2454559.18096$^{+0.00028}_{-0.00028}$\\
\noalign{\smallskip}
$T_{\rm 14}$ (d)
& 0.1822$^{+0.0019}_{-0.0023}$ 
& 0.1825$^{+0.0017}_{-0.0017}$
& 0.1824$^{+0.0016}_{-0.0016}$\\
\noalign{\smallskip}
$T_{\rm 12} = T_{\rm 34}$ (d)
& 0.0235$^{+0.0019}_{-0.0030}$ 
& 0.0236$^{+0.0017}_{-0.0018}$
& 0.0239$^{+0.0017}_{-0.0017}$\\
\noalign{\smallskip}
$\Delta F = R_{\rm P}^{2}$/R$_{*}^{2}$ 
& 0.01672$^{+0.00029}_{-0.00035}$ 
& 0.01674$^{+0.00027}_{-0.00027}$ 
& 0.01678$^{+0.00026}_{-0.00027}$\\
\noalign{\smallskip}
$b$ $\equiv$ $a \cos i/R_{\rm *}$ 
& 0.352$^{+0.075}_{-0.316}$ 
& 0.355$^{+0.068}_{-0.111}$ 
& 0.370$^{+0.064}_{-0.096}$\vspace{3 mm}\\
\noalign{\smallskip}
$K_{\rm 1}$ (\kms)
& 0.0569$^{+0.0055}_{-0.0053}$ 
& 0.0592$^{+0.0058}_{-0.0057}$
& 0.0564$^{+0.0051}_{-0.0051}$\\
\noalign{\smallskip}
$\gamma$ (\kms)
& -49.5128$^{+0.0016}_{-0.0016}$ 
& -49.5128$^{+0.0014}_{-0.0014}$
& -49.5125$^{+0.0014}_{-0.0015}$\\
\noalign{\smallskip}
$\gamma_{\rm HARPS}-\gamma_{\rm COR.}$ (\kms)
& 0.0267$^{+0.0034}_{-0.0035}$ 
& 0.0297$^{+0.0023}_{-0.0023}$
& 0.0233$^{+0.0015}_{-0.0015}$\\
\noalign{\smallskip}
$a$ (AU)
& 0.0501$^{+0.0017}_{-0.0018}$ 
& 0.0494$^{+0.0017}_{-0.0018}$
& 0.0507$^{+0.0017}_{-0.0018}$\vspace{3 mm}\\
\noalign{\smallskip}
$i$ (deg)
& 87.8$^{+2.0}_{-1.0}$ 
& 88.16$^{+0.58}_{-0.45}$
& 86.95$^{+0.87}_{-0.63}$\\
\noalign{\smallskip}
$e \cos \omega$ 
& 0.036$^{+0.034}_{-0.031}$ 
& 0.034$^{+0.025}_{-0.024}$
& \ldots \\
\noalign{\smallskip}
$e \sin \omega$ 
& $-0.10^{+0.13}_{-0.13}$
& $-0.233^{+0.071}_{-0.070}$
& \ldots \\
\noalign{\smallskip}
$e$ 
& 0.129$^{+0.106}_{-0.068}$ 
& 0.237$^{+0.068}_{-0.069}$
& 0. (fixed) \\
\noalign{\smallskip}
$\omega$ (deg)
& $290^{+106}_{-16}$ 
& $278.0^{+8.2}_{-5.6}$
& \ldots \\
\noalign{\smallskip}
$\phi_{\rm mid\mbox{-}eclipse}$
& $0.523^{+0.021}_{-0.020}$ 
& $0.522^{+0.016}_{-0.015}$
& 0.5 (fixed) \\
\noalign{\smallskip}
$T_{\rm 58}$ (d) $^{a}$
& 0.152$^{+0.040}_{-0.034}$ 
& 0.117$^{+0.017}_{-0.015}$
& 0.1824$^{+0.0016}_{-0.0016}$ (fixed)\\
\noalign{\smallskip}
$T_{\rm 56} = T_{\rm 78}$ (d) $^{b}$
& 0.0186$^{+0.0063}_{-0.0046}$ 
& 0.0140$^{+0.0022}_{-0.0019}$
& 0.0239$^{+0.0017}_{-0.0017}$ (fixed)\vspace{3 mm}\\
\noalign{\smallskip}
$\lambda$ (deg)
& $-147^{+49}_{-11}$ 
& $-148.7^{+13.7}_{-9.3}$
& $-149.3^{+11.5}_{-8.9}$\\
\noalign{\smallskip}
\vsini\ (\kms)
& 20.0$^{+69.2}_{-5.2}$ 
& 19.1$^{+5.7}_{-4.8}$
& 19.1$^{+5.3}_{-4.7}$\vspace{3 mm}\\
\noalign{\smallskip}
$M_{\rm *}$ (\msol)
& 1.20$^{+0.12}_{-0.12}$ 
& 1.16$^{+0.12}_{-0.12}$ 
& 1.25$^{+0.13}_{-0.13}$\\
\noalign{\smallskip}
$R_{\rm *}$ (\rsol)
& 1.38$^{+0.20}_{-0.18}$ 
& 1.200$^{+0.081}_{-0.080}$
& 1.566$^{+0.073}_{-0.073}$\\
\noalign{\smallskip}
$\log g_{*}$ (cgs)
& 4.23$^{+0.12}_{-0.12}$ 
& 4.341$^{+0.068}_{-0.068}$
& 4.143$^{+0.032}_{-0.031}$\\
\noalign{\smallskip}
$\rho_{\rm *}$ ($\rho_{\rm \odot}$)
& 0.45$^{+0.23}_{-0.15}$ 
& 0.67$^{+0.16}_{-0.13}$ 
& 0.323$^{+0.035}_{-0.028}$ \vspace{3 mm}\\
\noalign{\smallskip}
$M_{\rm P}$ (\mjup)
& 0.490$^{+0.059}_{-0.056}$ 
& 0.496$^{+0.064}_{-0.060}$
& 0.498$^{+0.059}_{-0.056}$\\
\noalign{\smallskip}
$R_{\rm P}$ (\rjup)
& 1.74$^{+0.26}_{-0.23}$ 
& 1.51$^{+0.10}_{-0.10}$
& 1.97$^{+0.10}_{-0.10}$ \\
\noalign{\smallskip}
$\log g_{\rm P}$ (cgs)
& 2.56$^{+0.14}_{-0.13}$ 
& 2.696$^{+0.086}_{-0.083}$
& 2.466$^{+0.051}_{-0.052}$\\
\noalign{\smallskip}
$\rho_{\rm P}$ ($\rho_{\rm J}$)
& 0.092$^{+0.054}_{-0.032}$ 
& 0.144$^{+0.042}_{-0.031}$
& 0.0648$^{+0.0106}_{-0.0090}$\\
\noalign{\smallskip}
$T_{\rm P,A=0}$ (K)
& 1662$^{+113}_{-110}$ 
& 1557$^{+55}_{-55}$
& 1756$^{+26}_{-30}$\\
\noalign{\smallskip}
\\ 
\hline 
\multicolumn{4}{l}{Three solutions are presented (Cases 1, 2 and 3), each with 
different constraints as described in the text (\S\ref{sec:sysparams}).}\\
\multicolumn{4}{l}{$^{a}$ $T_{\rm 58}$: total eclipse duration. 
$^{b}$ $T_{\rm 56} = T_{\rm 78}$: eclipse ingress/egress duration.}\\
\end{tabular}
\end{table*} 

\subsection{Transit times}
We measured the times of WASP-17b's transits to search for transit timing
variations, as may be induced by a third body (e.g., Holman \& Murray 2005; Agol
et al. 2005). 
The model light curves were stepped in time over the phometric data around the 
predicted times of transit, and $\chi^{2}$ was calculated at each step. 
The times of mid-transit were found by measuring the $\chi^{2}$ minima and the
uncertainties were determined via bootstrapping. 
The calculated times of mid-transit, $T_{\rm c}$, and the differences, $O-C$, 
between those times and the predicted times, assuming a fixed epoch and period 
(Table~\ref{sys-params2}), are given in Table~\ref{tabTTV}. 
No significant departure from a fixed ephemeris is seen. 


\begin{table}[!h]
\centering
\caption{Transit times\label{tabTTV}} 
\begin{tabular}{rrrr}
\hline
\hline
  $N_{tr}$ & $T_{\rm c}$~~~~~~~~ &   $\sigma_{T_{\rm c}}$~ & $O-C$ \\
           & (HJD)~~~~~       &   (min)                & (min) \\
\hline
$-$179 & 2453890.54850 & 6.2~\, & 16.6~\, \\
$-$175 & 2453905.48193 & 5.5~\, & 4.7~\, \\
$-$171 & 2453920.42243 & 3.6~\, & 2.8~\, \\
$-$159 & 2453965.23733 & 5.0~\, & $-$12.2~\, \\
$-$96  & 2454200.57081 & 4.4~\, & $-$11.2~\, \\
$-$92  & 2454215.52194 & 2.7~\, & 2.3~\, \\
$-$77  & 2454271.55701 & 4.1~\, & 7.2~\, \\
$-$73  & 2454286.49409 & 8.3~\, & 0.5~\, \\
$-$69  & 2454301.45133 & 8.2~\, & 22.8~\, \\
$-$61  & 2454331.32310 & 9.3~\, & 5.8~\, \\
$-$1   & 2454555.43662 & 6.4~\, & $-$12.9~\, \\
2      & 2454566.65046 & 8.3~\, & $-$2.0~\, \\
9      & 2454592.80049 & 0.55 & 0.79 \\
\hline
\end{tabular}
\end{table}

\newpage

\section{Discussion}
WASP-17b is the least dense planet known, with a density of 
0.06--0.14\,$\rho_{\rm J}$.
TrES-4, the previous least dense planet, has a density of 0.15\,$\rho_{\rm J}$ 
(Sozzetti et al. 2009), 
and HD\,209458b, the most studied transiting planet, has a density of 
0.27\,$\rho_{\rm J}$ (Torres et al. 2008).
WASP-17b's radius of 1.5-2\,\rjup\ is larger than predicted by standard planet 
evolution models.
For example those of Fortney et al. (2007) imply a radius of at most 1.3\,\rjup\
(the value for a 1-Gyr-old, coreless planet of 0.41\,\mjup, receiving more 
stellar flux, at a distance of 0.02\,AU, where each of these values errs on the 
side of inflating the radius).

Burrows et al. (2007) showed that an enhanced atmospheric opacity can delay
radius shrinkage, leading to a larger-than-otherwise planet radius at a given
age. Enhanced opacities may result from super-solar metallicity, the presence of
clouds/hazes, or the effects of photolysis or non-equilibrium chemisty. 
One might expect a low planetary atmospheric opacity due to the sub-solar 
metallicity of the WASP-17 star: [Fe/H] = --0.25. However, as WASP-17b is 
highly irradiated, its atmospheric opacity is expected to be high due, for
example, to the presence of TiO and VO gases (Burrows et al. 2008; Fortney 
2008).
Ibgui \& Burrows (2009) 
found that an atmospheric opacity of $3 \times {\rm solar}$ is sufficient
to account for the radius of HD\,209458b.
The very large radius of WASP-17b and its moderate age suggest 
that enhanced opacity alone, even of $10 \times {\rm solar}$, is insufficient to
account for the planet's bloatedness. 

It has been proposed (Bodenheimer et al. 2001; Gu et al. 2003; Jackson et al. 
2008a) that tidal dissipation associated with the circularisation
of an eccentric orbit is able to substantially inflate the radius of a 
short-orbit, giant planet.
If a planet is in a close ($a < 0.2$ AU), highly eccentric ($e > 0.2$) 
orbit then planetary tidal dissipation will be significant and will shorten and 
circularise the orbit. Orbital energy is deposited within the 
planet interior, leading to an inflated planet radius. 
This process is accelerated by higher atmospheric opacities: as the planet
better retains heat, shrinking of the radius is retarded, and a larger radius
causes greater tidal dissipation.
Higher eccentricities result in stronger tides, and thus in greater tidal
dissipation.
The rate at which energy is tidally dissipated within a body is 
inversely proportional to its tidal quality factor, $Q'$, which is the 
ratio of the energy in the tide to the tidal energy dissipated within the body 
per orbit (e.g., Ogilvie \& Lin 2007).

Ibgui \& Burrows (2009) created a tidal dissipation model and 
applied it to HD\,209458b, which is bloated to a lesser degree than WASP-17b.
A custom fit is required to find possible evolution histories for the WASP-17 
system, but the similarity of HD\,209458 
(compare Tables~\ref{wasp17-params} and \ref{sys-params2} from this 
paper with Table~1 in Ibgui \& Burrows (2009) and references therein) permits 
comparison.
%
Ibgui \& Burrows' (2009) HD\,209458b model suggests that tidal heating could 
produce even WASP-17b's maximum likely radius ($R_{\rm P} \approx 2\,\rjup$)  
if, for example, it evolved from a highly eccentric ($e \approx 0.79$), close 
($a \approx 0.085$ AU)
orbit, with moderate tidal dissipation ($Q'_{\rm P} \approx 10^{6.55}$; 
$Q'_{\ast} \approx 10^{7.0}$) and solar atmospheric opacity. 
The final semimajor axis of this particular model was shorter than that of 
WASP-17b, but within 10 per cent.
Such an eccentric, short orbit seems reasonable as
planets in highly eccentric, quite short orbits are known: 
HD\,17156b ($e = 0.676$, $a = 0.162$ AU; Winn et al. 2009d), 
HD\,37605b ($e = 0.74$, $a = 0.26$ AU; Cochran et al. 2004), 
HD\,80606b ($e = 0.934$, $a = 0.45$ AU; Moutou et al. 2009).
As for the strength of tidal dissipation, Jackson et al. (2008b) found similar
best-fitting values ($Q'_{\rm P} = 10^{6.5}$; $Q'_{\ast} = 10^{5.5}$) 
when matching the current eccentricities of short-orbit ($a < 0.2$ AU) planets 
with the eccentricities of farther out planets, from which they presumably 
evolved.

The limited radial-velocity measurements during transit give a
strong indication that WASP-17b is in a retrograde orbit.
As the angular momenta of a star, its protoplanetary disc, and hence its 
planets, all derive from that of the parent molecular cloud, WASP-17b presumably
originated in a prograde orbit.
As a gas giant, WASP-17b is expected to have formed just outside the ice 
boundary ($\sim$3\,AU) and migrated inwards to its current separation of 
0.05\,AU (e.g., Ida \& Lin 2004).
Migration of a giant planet via tidal interaction with a gas disc is expected 
to preserve spin-orbit alignment (Lin et al. 1996; Ward 1997) and is thus unable
to produce a retrograde orbit.
Alternatively, migration via a combination of star-planet scattering (Takeda et 
al. 2008) or planet-planet scattering (e.g., Rasio \& Ford 1996; Chatterjee et 
al. 2008; Nagasawa  et al. 2008) and tidal circularisation of the resultant 
eccentric orbit is able to produce significant misalignment.

In addition to inclined orbits, scattering is able to produce highly 
eccentric orbits (e.g., Ford \& Rasio 2008), which have been found to be common 
and are necessary if tidal circularisation is to inflate planetary radii
(e.g., Jackson et al. 2008b), whereas planet-disc interactions seem unable to 
pump eccentricities to large values ($e > 0.3$; D'Angelo et al. 2006).
Nagasawa et al. (2008) carried out orbital integrations of three-planet systems:
three Jupiter-mass planets were initially placed beyond the ice boundary 
(5\,AU, 7.25\,AU, 9.5\,AU) in circular orbits, with small inclinations 
($0.5^{\circ}$, $1^{\circ}$, $1.5^{\circ}$), around a solar-mass star.
They found that a combination of planet-planet scattering, the Kozai mechanism 
(the oscillation of the eccenticity and inclination of a planet's orbit via the
secular perturbation from outer bodies; Kozai 1962)
and tidal circularisation produces short-orbit, giant planets in $\sim$30\% of 
cases. The Kozai mechanism is most effective when the scattered, inner
planet has an inclined orbit.
A broad spread was seen in the inclination distribution of the short-orbit
planets formed, including planets in retrograde orbits.
Therefore, we suggest that WASP-17b supports the hypothesis  
that some short-orbit, giant planets are produced by a combination of 
scattering, the Kozai mechanism and tidal circularisation. 
The observation of the RM effect for more short-orbit planets is 
required to measure the size of the contribution. 

For planet-planet or star-planet scattering to have influenced WASP-17b's orbit 
in the past, one or more stellar or planetary companions must have been present 
in the system. 
Sensitive imaging could probe for a stellar companion and further radial 
velocity measurements are necessary to search for stellar or planetary 
companions. 
It will be worthwhile looking for long-term trends in the RVs to detect
farther-out planets that might have been involved in past scattering.
A straight-line fit to the residuals of the RV data about the model fits
indicates no significant drift over a span of 622\,days 
(e.g., for Case 1 the drift is $-17 \pm 11 \,\ms$).
In their three-planet integrations, Nagasawa et al. (2008) found that in 75\% of
cases one planet is ejected, a planet collides with the host star in 22\% of 
cases, and two planets are ejected in 5\% of cases.
They also found that, since a small difference in orbital energy causes a 
large difference in semimajor axis in the outer region, the final semimajor 
axes of outer planets are widely distributed (peak at $\sim$15\,AU, with a large
spread).
Therefore, it is possible that WASP-17 is now the only giant planet in the 
system or that the outer planets are in long orbits, which are difficult to 
detect with the RV technique.

\begin{figure}
\plotone{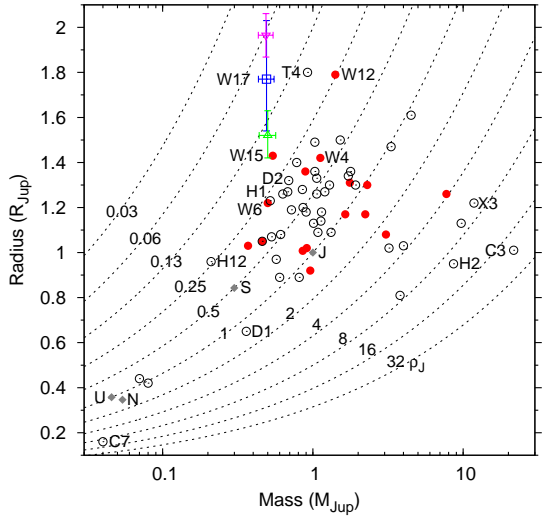}
\caption{Mass-radius distribution of the 62 known transiting extrasolar planets. 
The best-fitting values for the three WASP-17b models are depicted according to
the key given in Table~\ref{sys-params2}. 
Other WASP planets are filled, red circles; 
non-WASP planets are open, black circles; 
Jupiter, Saturn, Neptune and Uranus are filled, grey diamonds, 
labelled with the planets' initials. 
For clarity, error bars are displayed only for WASP-17b. 
Some planets discovered by CoRoT, HAT, TrES, WASP and XO are labelled with the
project initial and the system number (e.g., {\bf W}ASP-{\bf 17}b = {\bf W17}). 
HD\,149026b is labelled D1 and HD\,209458b is labelled D2.
The labelled, dashed lines depict a range of density contours in Jovian units.
Data are taken from this work and http://exoplanet.eu.
\label{figMassRadius}}
\end{figure}

The discovery of WASP-17b extends the mass-radius distribution 
of the 62 known transiting exoplanets (Figure~\ref{figMassRadius}). 
WASP-17b has the largest atmospheric scale height (1100--2100 km) of any 
known planet by up to a factor 2, due to its very low surface gravity and 
moderately high equilibrium temperature. 
The ratio of projected areas of planetary atmosphere to stellar 
disc of WASP-17b is 1.9--2.7 times that of HD\,209458b and
2.4--3.4 times that of HD\,189733b, for both of which successful 
attempts at measuring atmospheric signatures have been made 
(e.g., Charbonneau et al. 2002, Desert et al. 2009). 
Thus, although WASP-17 is fainter and has a larger stellar
radius, the system is a good prospect for transmission spectroscopy.

\section{Acknowledgements}
We acknowledge a thorough and constructive report from the anonymous referee. 
The WASP consortium comprises the Universities of Keele, Leicester, St. Andrews, 
the Queen's University Belfast, the Open University and the Isaac Newton Group. 
WASP-South is hosted by the South African Astronomical Observatory and we are 
grateful for their support and assistance. 
Funding for WASP comes from consortium universities and from the UK's Science 
and Technology Facilities Council.


\end{document}